\documentclass[twocolumn,amsmath,amssymb,pra,superscriptaddress]{revtex4}

\usepackage{graphics}
\usepackage{epsfig}
\usepackage{amsmath}
\usepackage{amsfonts}
\usepackage{color}
\usepackage{dcolumn}
\usepackage{bm}
\usepackage{mathrsfs}
\usepackage{float}

\begin{document}

  \bibliographystyle{apsrev}
  
\title{Comparison of the spatial-phase retardation effects between Rayleigh and Taylor plane-wave multipole series approximations}

      \author{Eric Ouma Jobunga}
       \affiliation{\it  Department  of Mathematics and Physics , Technical University   of Mombasa,\\ P. O. Box 90420-80100, Mombasa, Kenya}
         
%



\begin{abstract}

As experimentalists explore new opportunities in strong field radiation offered by current generation light sources,
new theoretical tools become inevitable in dealing with the challenging non-linear dynamics that come into play as a result of the increasing laser intensities and the shorter wavelengths. While many theoretical
studies employ the electric dipole approximation for convenience reasons, in the strong-field regime the validity of this 
approximation is questionable. We have made a detailed comparison of the expansion of the retardation term,
$e^{i\mathbf{k} \cdot \mathbf{r}}$ in both Taylor and Rayleigh series multipole approximations
with the angle between the radial vector and the direction of propagation chosen arbitrarily to be $45^{\circ}$.
It is verified in this paper that the Rayleigh plane-wave expansion provides a larger validity
range in comparison to the widely used Taylor expansion. We also take note that the Taylor approximated spherical Bessel functions reproduce the lower limits of the regular spherical Bessel functions but deviates strongly in the asymptotic region. We conclude that the use of
the Rayleigh plane-wave expansion provides the most accurate contribution of any given order of the multipole expansion. The discrepancy in the dipole and non-dipole photoelectron energy spectra as predicted by these approximations using short-wavelength intense laser pulses interacting with hydrogen atom in its ground state show the importance of the higher-order terms absent in the Taylor expansion.

\end{abstract}

\maketitle

\section{Introduction}
\label{sec:Intro}
 The development of the free-electron lasers and new generation light sources \cite{amo:Couprie2014} has enabled the realisation of high precision experiments 
 investigating various non-linear processes in the dynamics of atomic, molecular, and ionic systems
 interacting with laser pulses whose intensities and duration are in the order of $\sim 10^{23}$ W cm$^{-2}$ and attosecond 
 time scale respectively \cite{Scrinzi2006}. The analysis of such experiments definitely
 require reliable non-perturbative solution of the time dependent Schr\"{o}dinger or Dirac equation. These solutions
 should consider both temporal and spatial intensity variations of the laser pulse \cite{Freeman1991}.
 
 While most strong-field theoretical studies have been concentrated on the spatially independent electric dipole approximation, there have
 been some previous attempts to incorporate the non-dipole effects \cite{Amusia1999,Kylstra2000, amo:Demekhin2014,Valery2001,Forre2006,Zhongyuan2013,Dondera2012,Varma2009,Viorica2011, Forre2014} which come into play when the spatial variation of the vector potential is considered. The results of these non-dipole effects have
 predicted certain effects like total breakdown of the electric dipole approximation for hard x-rays \cite{amo:Demekhin2014,Valery2001}, new structures in the photoelectron angular distribution and distortion of the dipole photoelectron energy spectra in the XUV and soft x-ray wavelengths with intensities greater than unity (in atomic units) \cite{Forre2006,Zhongyuan2013}, small distortion of the dipolar angular distribution for very small wavelengths and intensities of $\sim 1$ a.u. \cite{Dondera2012}, and the breakdown of stabilization with intense high frequency laser pulses\cite{Kylstra2000}. From a theoretical point of view, the contribution of the non-dipole effects arising from the $\mathbf{A}^2$ term  relative  to the $\mathbf{A} \cdot \mathbf{p}$ term in the strong-field regime has been a subject of conflicting viewpoints with some literature attributing the $\mathbf{A}^2$ term non-dipole corrections to be dominant \cite{Forre2006, Forre2014,Varma2009} and the others attributing the $\mathbf{A} \cdot \mathbf{p}$ non-dipole corrections to be dominant \cite{Viorica2011,Dondera2012,Zhongyuan2013}. Although diverse theoretical
 approaches have been used to analyze these
 strong-field effects, one of the ingredients that has been apparently employed in common is the use of the Taylor approximation of the spatial phase 
 retardation term, $e^{i\mathbf{k} \cdot \mathbf{r}}$, to include the non-dipole effects \cite{amo:BetheandSalpeter1957}.  But in the plane-wave description of electromagnetic radiation, despite being usual, it is not essential to expand the electric field in Taylor series \cite{Craig1998}. The expansion of electromagnetic plane-waves in terms of Bessel functions and spherical harmonics dates back to the theoretical work of the $1930s$ \cite{amo:Roger2008, Heitler1936} and this expansion in twisted beams allows a very direct connection to be made between the angular momentum of the photon and the terms of the expansion.

 In this paper, we compare the expansion of the spatial phase retardation term using the regular spherical Bessel functions (SBA)\cite{cp:Riley2006}, also known as the Rayleigh plane-wave expansion,
 and the corresponding Taylor plane-wave multipole expansion (TA) approximations. The main difference between the two approaches stems from the use of spherical Bessel functions in the Rayleigh expansion and the use of polynomials in the Taylor expansion. Our strategy in comparing the two parallel approaches involves rewriting the Taylor expanded terms of the retardation into Legendre polynomials. From 
 the coefficients of the Legendre polynomials we obtain unique analytical functions which are correlated with each of the regular 
 spherical Bessel functions. We then use these analytical functions and the alternative regular spherical Bessel functions to model the retardation term. We further analyze the effect of the two alternative methods in the photoelectron energy (PE) spectrum of a strongly driven hydrogen atom. We consider the interactions up to the hexadecapole term of the interaction Hamiltonian but focussing only on the $\mathbf{A} \cdot \mathbf{p}$  interaction. A comprehensive treatment of the non-dipole effects would be considered in a subsequent paper.
 
 \section{Theory}
 \label{sec:theo}
The non-relativistic dynamics of atoms interacting with a classical electromagnetic field is governed by the 
 time-dependent Schr\"{o}dinger equation
   \begin{equation}
    i \frac{\partial \Psi(\mathbf{r},t)}{\partial t} = [\mathrm{H}_0 + \mathrm{V}(\mathbf{r},t)] \Psi(\mathbf{r},t) \label{eq:1}
   \end{equation}
where $\mathrm{H}_0$ is the unperturbed Hamiltonian corresponding to the field-free eigenstates, 
$\mathrm{V}(\mathbf{r},t)$ 
    \begin{equation}
     \mathrm{V}(\mathbf{r},t) = -q \mathbf{A} \cdot \mathbf{p} + \frac{1}{2} q^2\mathbf{A}^2            \label{eq:2}
    \end{equation}
is the radiation gauge interaction potential expressed in terms of momentum operator $\mathbf{p}$ and
the vector potential $\mathbf{A}(\mathbf{r},t)$, with $q$ as the electronic charge.    
  The vector potential satisfies the Coulomb gauge condition, $\nabla \cdot
  \mathbf{A}(\mathbf{r},t) = 0 $. In this problem, a linearly polarised pulse with the vector potential
   \begin{equation}
    \mathbf{A}(\mathbf{r},t) \approx A_0 f(t) \mathrm{sin}{(\mathbf{k} \cdot \mathbf{r}-\omega t + \delta)} \hat{z}
   \end{equation}
   in the $+$z direction and the wave vector $\mathbf{k}$ oriented in the $+\hat{x}$ direction is considered. Here,
   $A_0 = E_0/\omega$ is the amplitude of the vector potential, $E_0$ is the peak electric field strength, $\omega$ defines the laser frequency, $f(t)$
   is the laser  pulse carrier envelope function with $\delta$ as the carrier envelope phase. The spatial dependence of the carrier envelope function is assumed to be ignorable. The 
   envelope function can be freely chosen as a $\cos^2$, Gaussian, or any other pulse shape but with a periodic pulse that is only non-zero when 
   the time $t$ is enclosed within the set $(0,\tau)$ with $\tau = (2 \pi N/ \omega)$ as the total pulse duration for a laser pulse containing $N$ photons. 
This expression for the vector potential  can be expanded as 
   \begin{equation}
    \mathbf{A}(\mathbf{r},t) \approx A_0 f(t) [\mathrm{sin}(\mathbf{k} \cdot \mathbf{r}) \cos(\omega t ) - 
                         \cos(\mathbf{k} \cdot \mathbf{r}) \mathrm{sin}(\omega t )] \hat{z}
   \end{equation}
   with the phase angle $\delta = 0$ chosen for the sake of convenience. The spatial terms can  then be expressed in terms exponential (retardation) term $e^{i\mathbf{k} \cdot \mathbf{r}}$ 
     \begin{equation}
     \begin{split}
      \cos (\mathbf{k} \cdot \mathbf{r}) &= \frac{1}{2}[e^{i\mathbf{k} \cdot \mathbf{r}} + e^{-i\mathbf{k} \cdot \mathbf{r}}]\\
      \mathrm{sin} (\mathbf{k} \cdot \mathbf{r}) &= \frac{1}{2i}[e^{i\mathbf{k} \cdot \mathbf{r}} - e^{-i\mathbf{k} \cdot \mathbf{r}}]
     \end{split} 
     \end{equation}
 and its conjugate.    
The retardation term can subsequently be expanded in the Taylor series
     \begin{equation}
      e^{i\mathbf{k} \cdot \mathbf{r}} = 1 + (i\mathbf{k} \cdot \mathbf{r}) + \frac{(i\mathbf{k} \cdot \mathbf{r})^2}{2!} + \cdots + \frac{(i\mathbf{k} \cdot \mathbf{r})^l}{l!}  \label{eq:t1}
     \end{equation}
 or equivalently in terms of the well known Rayleigh multipole expansion series which employ the regular spherical Bessel functions $j_l(kr)$
\cite{amo:Bransden1990,amo:Devanathan2002} and  the spherical harmonics \cite{amo:Roger2008} 
       \begin{equation}
      e^{i\mathbf{k} \cdot \mathbf{r}} = 4 \pi \sum_l^{\infty} \sum_{m=-l}^{+l} i^l j_l(kr) 
                                    Y_{l,m}^*(\hat{\mathbf{k}})Y_{l,m}(\hat{\mathbf{r}})  \label{eq:j2}
      \end{equation}
with the order of the multipole expansion defined by $l=0,1,2,3, \cdots $ as the dipole, quadrupole, octupole, hexadecapole, and other higher multipole-order  terms.

To compare the two expansions, we consider only the first six terms of the retardation expansion in Taylor series 
    \begin{equation}
    \begin{split}
     e^{i\mathbf{k} \cdot \mathbf{r}} &= 1 + ikr\cos\theta - \frac{k^2r^2\cos^2\theta}{2} -i \frac{k^3r^3\cos^3\theta}{6}\\
                                          &+ \frac{k^4r^4\cos^4\theta}{24} + i \frac{k^5r^5\cos^5\theta}{120} + \cdots
    \end{split}
    \end{equation}
and express them in terms of the Legendre polynomials $P_l(\cos \theta)$  \cite{Boas2006}
    \begin{equation}
    \begin{split}
     e^{i\mathbf{k} \cdot \mathbf{r}} &= P_0(\cos\theta) + ikrP_1(\cos\theta)\\&-\frac{k^2r^2}{6}[ P_0(\cos\theta) + 2P_2(\cos\theta)]\\
                                        &-\frac{ik^3r^3}{30}[3P_1(\cos\theta) + 2 P_3(\cos\theta)]\\ &+\frac{k^4r^4}{24 \times 35} [7P_0(\cos\theta) + 20P_2(\cos\theta) + 8P_4(\cos\theta)]\\ &+ \frac{ik^5r^5}{120 \times 63}[27P_1(\cos\theta) + 28P_3(\cos\theta) + 8P_5(\cos\theta)]                        
     \end{split}                                  
    \end{equation}
in order to take a similar form as equation (\ref{eq:j2}) for comparison convenience.   By arranging the terms in the orders of the Legendre polynomials, the retardation term can then be written as
      \begin{equation}
       e^{i\mathbf{k} \cdot \mathbf{r}} = \sum_l i^l\, h_l(kr)\, P_l(\cos\theta)     \label{eq:j3}              
      \end{equation}
where $h_l(kr)$ 
     \begin{equation}
      h_l(kr) = \frac{(kr)^l}{a_l} - \frac{(kr)^{l+2}}{2a_l(2l+3)} + \frac{(kr)^{l+4}}{8a_l(2l+3)(2l+5)} + \cdots
     \end{equation}
is  a polynomial with increasing orders of $kr$. Here $a_l$ is an $l$-dependent recursive term defined by $a_{l+1}=(2l+1)a_{l}$ relation and $a_0 = 1$ as the first term. Using these functions, the corresponding spherical Bessel functions
can then be constructed using the partial-wave decomposition,
    \begin{equation}
      \tilde{j}_l(kr) = \frac{1}{2l+1}h_l(kr) \label{eq:jh}
    \end{equation}

The recursive function $a_l$ can be shown to be equal to $(2l-1)!!$. If substituted in the above equation, we obtain the well known
expansion of the spherical Bessel functions \cite{Weisstein, Abramowitz1965}
    \begin{equation}
     \lim_{l_{\mathrm{max}\to \infty}} \tilde{j}_l^{(l_{\mathrm{max}})}(kr) = (kr)^l \sum_{n=0}^{n_{\mathrm{max}}}\,\frac{(-1)^n\, (kr)^{2n}}{2^n\,n!\, (2n+2l+1)!!}   \label{eq:jlh1}
    \end{equation}
    
with the integer $l_{\mathrm{max}}$ specifying the maximum multipole-order, $n_{\mathrm{max}}$ specifying the maximum degree of the summation of the $h_l(kr)$ functions, and $l$ denotes the order of the spherical Bessel function considered.

The degree $n_{\mathrm{max}}$ of the Taylor approximated spherical Bessel functions ($\tilde{j}_l^{(l_{\mathrm{max}})}$) specifies the number of terms to be included in the Taylor approximated series of the spherical Bessel function of order $l$. That is, the maximum number of terms $n_{\mathrm{max}}$ is defined in terms of the maximum order $l_{\mathrm{max}}$ by ${ n_{\mathrm{max}}= \mathrm{tr}\{(l_{\mathrm{max}}-l)/2\}}$ with $n_{\mathrm{max}}$ taking only the truncated integral values. In the Taylor approximated spherical Bessel functions, the degree varies with $l$ and $n$ as already shown in the relation.

The  functions $\tilde{j}_l^{(l_{\mathrm{max})}}(kr)$ can then be used to express the retardation
term $e^{i\mathbf{k} \cdot \mathbf{r}}$ in three dimensions using spherical harmonics expansion as
      \begin{equation}
       e^{i\mathbf{k} \cdot \mathbf{r}} = 4 \pi \sum_l^{\infty} \sum_{m=-l}^{+l}\, i^l\,\tilde{j}_l^{(l_{\mathrm{max})}}(kr) 
                                     Y_{l,m}^*(\hat{\mathbf{k}})Y_{l,m}(\hat{\mathbf{r}})     \label{eq:jlh}
      \end{equation}
 having made use of the known spherical harmonics \cite{amo:Devanathan2002} relation
      \begin{equation}
       \sum_{m=-l}^{+l}Y_{l,m}^*(\hat{\mathbf{k}})Y_{l,m}(\hat{\mathbf{r}}) = \frac{(2l+1)\mathrm{P}_l(\cos (\theta))}{4\pi}
      \end{equation}
  in equation (\ref{eq:j3})    with $\theta$ is the angle between the vectors $\mathbf{k}$ and $\mathbf{r}$  and $\mathrm{P}_l$ as the $l^{\mathrm{th}}$ order
      Legendre polynomial,    
This retardation expansion is similar to the multipole expansion using the regular spherical Bessel functions $j_l(kr)$ in equation (\ref{eq:j2})
       but different in the corresponding spherical Bessel functions $\tilde{j}_l^{(l_{\mathrm{max})}}(kr)$. The regular spherical Bessel functions $j_l(kr)$ are related to the ordinary Bessel functions of the first kind $J_m(kr)$ \cite{cp:Riley2006} by the rule 
     \begin{equation}
      j_l(kr) = \sqrt{\frac{\pi}{2kr}}J_{l+\frac{1}{2}}(kr)   \label{eq:j}
     \end{equation}
 with $m = l + 1/2$ as the order of the ordinary Bessel function.
 
 In making a useful comparison between the two different approaches in simulating non-linear dynamics, we can start by comparing directly the correlation between $j_l(kr)$ and $\tilde{j}_l^{(l_{\mathrm{max})}}(kr)$, and then subsequently comparing the components of the
retardation term by evaluating the trigonometric functions, ${\cos (\mathbf{k}\cdot \mathbf{r})}$ 
      and ${\mathrm{sin} (\mathbf{k}\cdot \mathbf{r})}$, as
     \begin{equation}
     \begin{split}
      \cos (\mathbf{k}\cdot \mathbf{r}) &= \sum_{l=0,2,\cdots,even}\, i^l\, j'_l(kr)\,(2l+1)\mathrm{P}_l(\cos (\theta))\\
      \mathrm{sin} (\mathbf{k}\cdot \mathbf{r}) &= \sum_{l=1,3,\cdots,odd}\, i^{l-1}\, j'_l(kr)\, (2l+1)\mathrm{P}_l(\cos (\theta))  \label{eq:trig}
     \end{split} 
     \end{equation}
 with $j'_l = j_l$ for the regular spherical Bessel functions and $j'_l = \tilde{j}_l^{(l_{\mathrm{max})}}$ for the $n^{th}$ degree Taylor approximated spherical Bessel functions.
 \begin{figure}[H]
  \centering
    \includegraphics*[width= 0.49\linewidth]{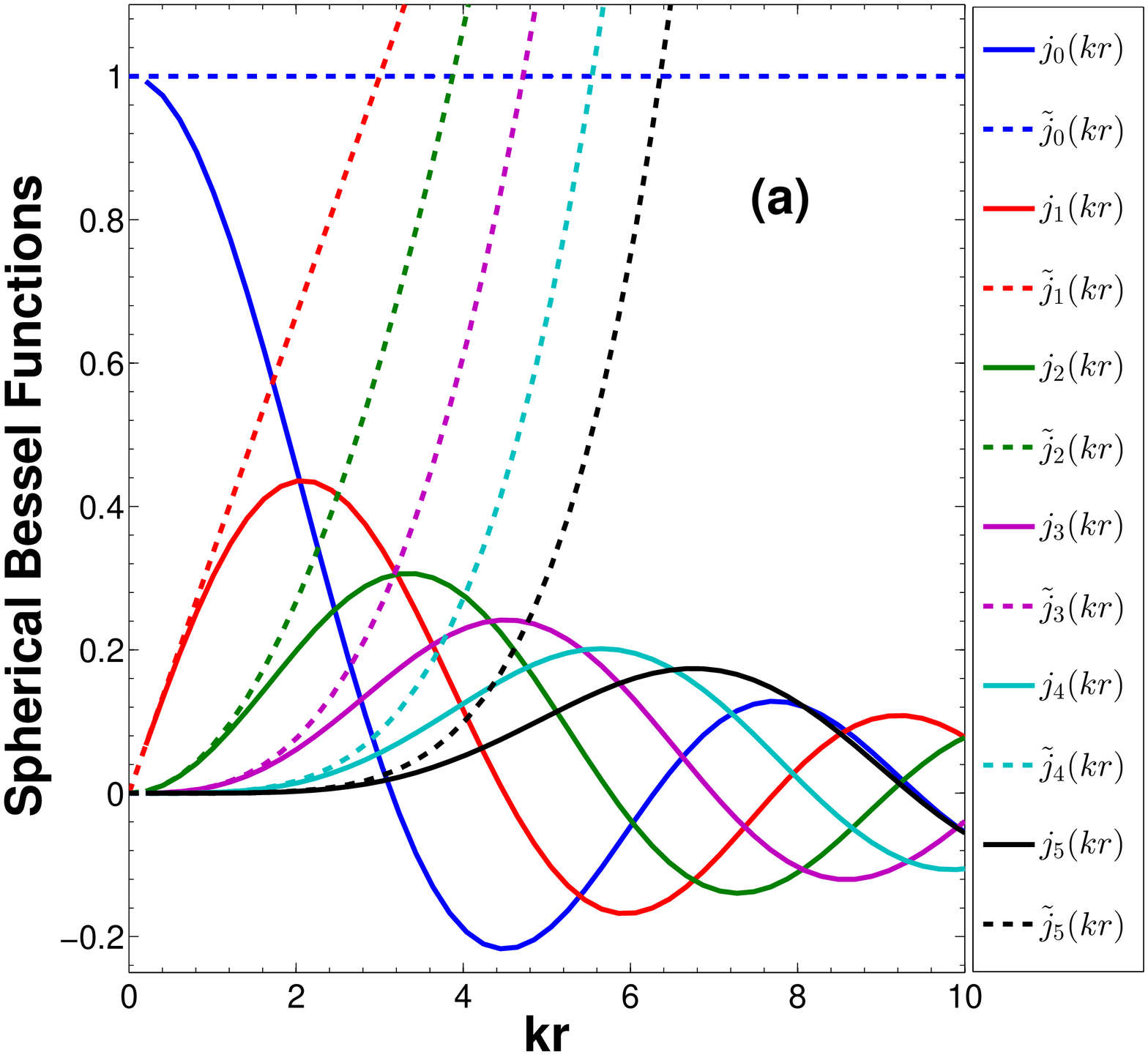}
  \includegraphics*[width= 0.47\linewidth]{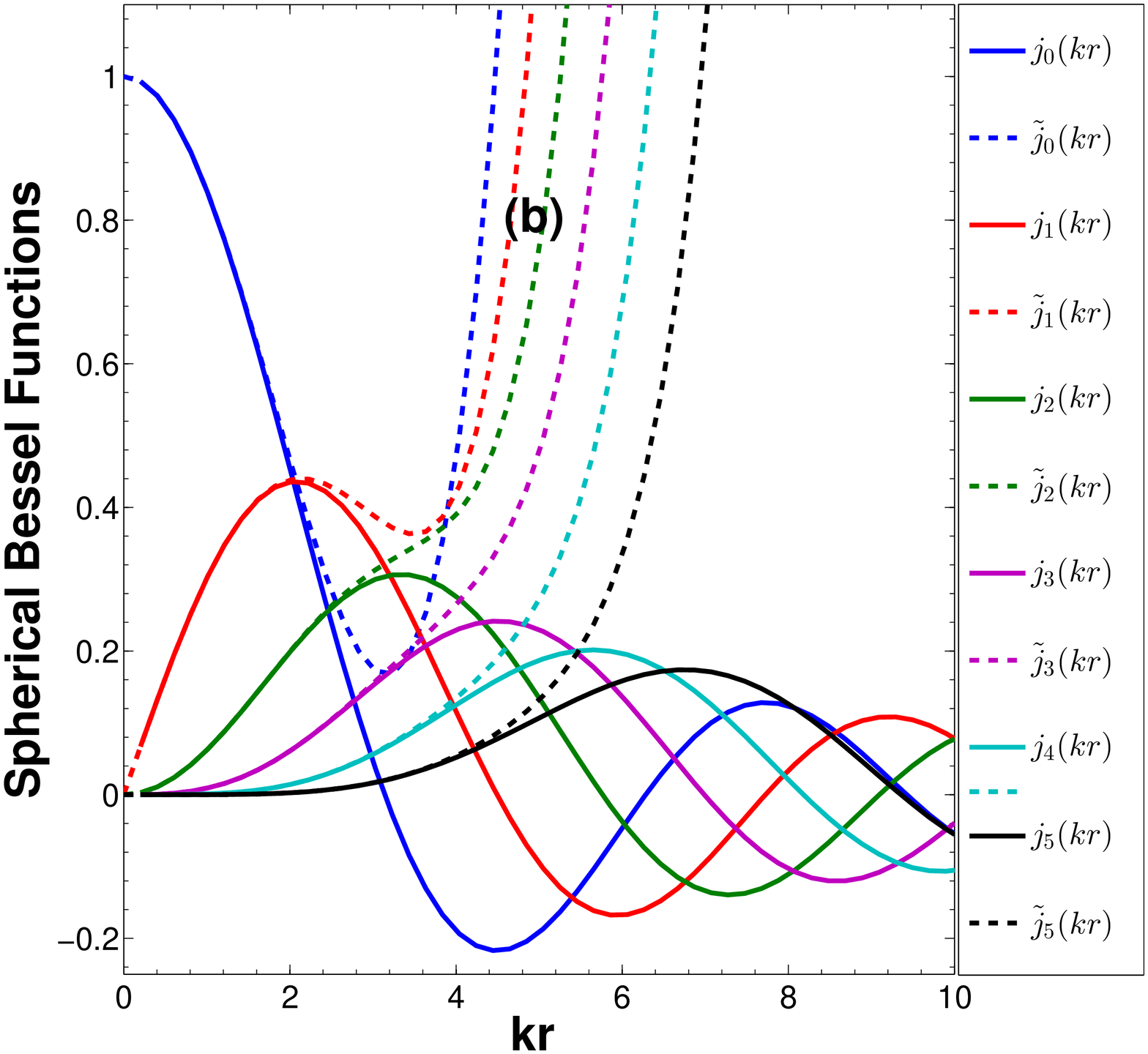}  
    \caption{(Color online)  Taylor approximated Bessel functions $\tilde{j}_l(kr)$ (broken lines) in comparison with the spherical Bessel functions $j_l(kr)$ (solid lines).
    Blue are $0^{\mathrm{th}}$ order, red are $1^{\mathrm{st}}$ order,
    green are $2^{\mathrm{nd}}$ order, violet are $3^{\mathrm{rd}}$ order, cyan are the $4^{\mathrm{th}}$ order, and black are the $5^{\mathrm{th}}$ order. Taylor approximated spherical Bessel functions $\tilde{j}_l(kr)$ is constructed from the (a) zero degree and (b) second degree approximation of the $h_l(kr)$ functions. }
  \label{JNfig}
\end{figure}

\section{Results and Discussion} 
\label{sec:res}
 In figure \ref{JNfig}, we have plotted the graphs of the first six of the Bessel functions, $\tilde{j}_l(kr)$, defined
 in equation (\ref{eq:jlh}), and the corresponding regular spherical Bessel functions, $j_l(kr)$.  The difference between the two figures is that in figure \ref{JNfig}(a) the summation in equation(\ref{eq:jlh1}) is truncated after the first term while in figure \ref{JNfig}(b), the summation is truncated after three terms.   While both Bessel functions are in excellent agreement in the 
 limit $kr \rightarrow 0$, certain key differences exist.
 First, it can be seen that the regular spherical Bessel functions are in general convergent towards a zero value as $kr \rightarrow \infty$  making the functions integrable whereas the functions $\tilde{j}_l(kr)$ are divergent and hence non-integrable beyond a certain critical point in $kr$. This could already be an indication of a possible breakdown in the Taylor
 approximation \cite{Strogatz2000} of a given order beyond the critical point in $kr$. Second, the Taylor approximated spherical Bessel functions approach the limit defined by the regular spherical Bessels of comparable order as the degree of approximation is increased. This basically means that if higher multipole-order terms are used in the Taylor expansion, the agreement between the two approximations become better and the validity regime of the Taylor approximation increases. 
 
 \begin{figure}[H]
  \center
    \includegraphics*[width= 0.49\linewidth]{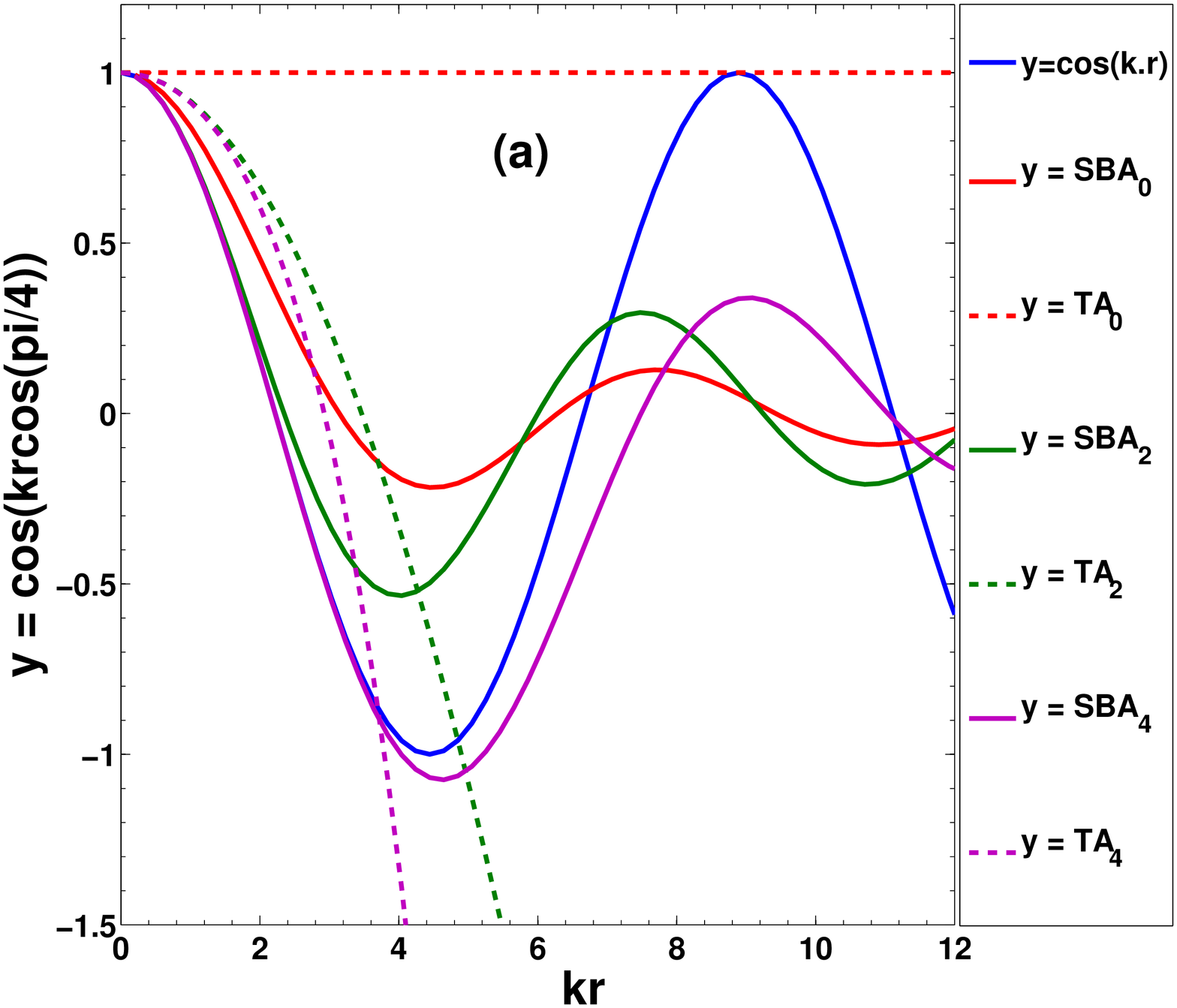}
    \includegraphics*[width= 0.49\linewidth]{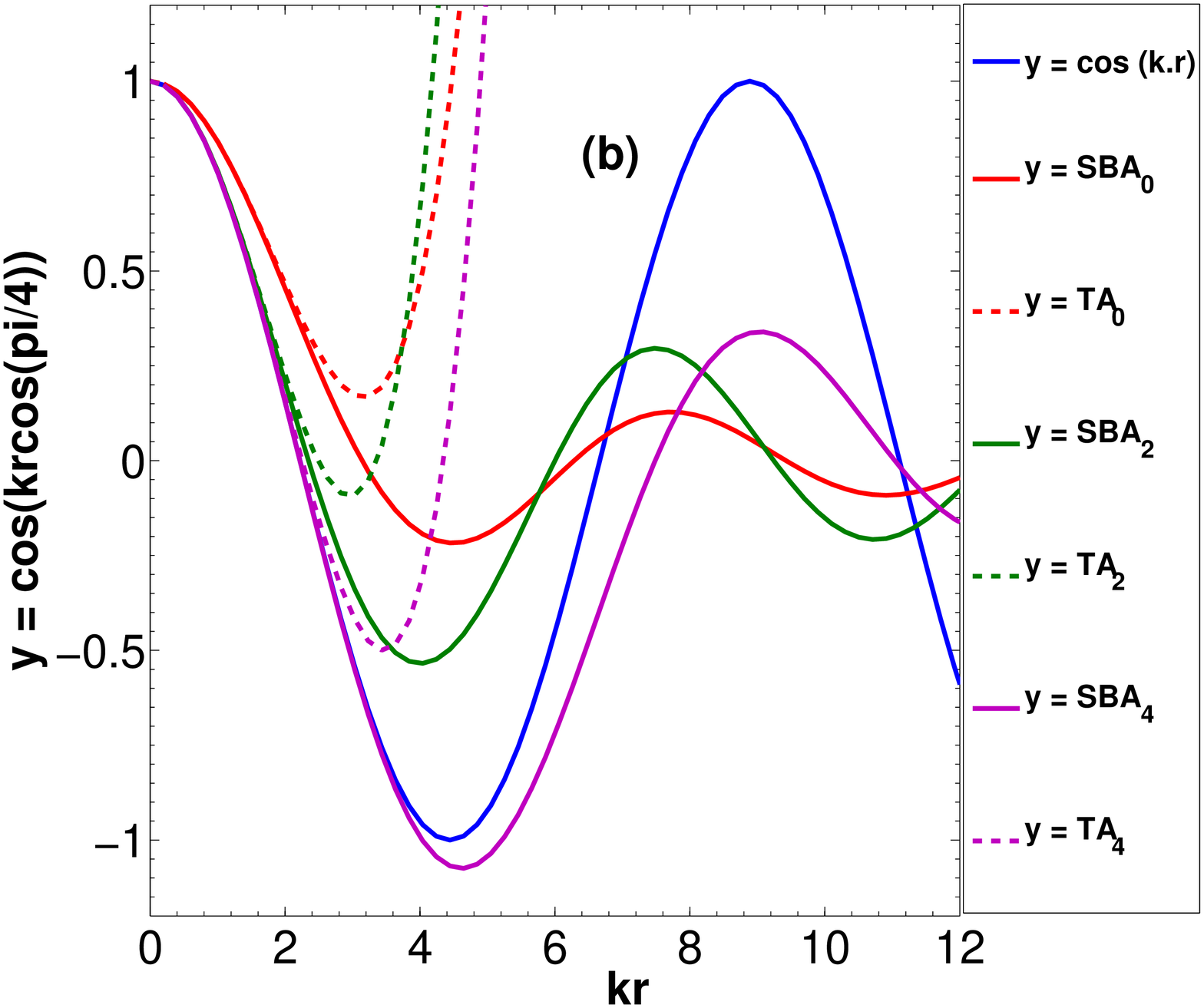}
    
    \caption{(Color online) The approximation of the cosine function ${y=\cos (\mathbf{k}\cdot \mathbf{r})}$ for the first six orders of
    the multipole expansion using the Taylor approximated spherical Bessel and the regular spherical Bessel functions expansion.  Blue curve is the ideal cosine function, while red, green, and violet are the $1^{\mathrm{st}}$, $2^{\mathrm{nd}}$,
     and $3^{\mathrm{rd}}$ order approximations respectively. Solid lines are the regular spherical Bessel function approximation while broken lines are the corresponding Taylor approximated spherical Bessel functions.
     A one-term Taylor approximation is used in (a) while (b) contains a summation of up to three terms in the Taylor approximation respectively.} \label{fig:ca}
\end{figure}

 \begin{figure}[H]
  \center
    \includegraphics*[width= 0.49\linewidth]{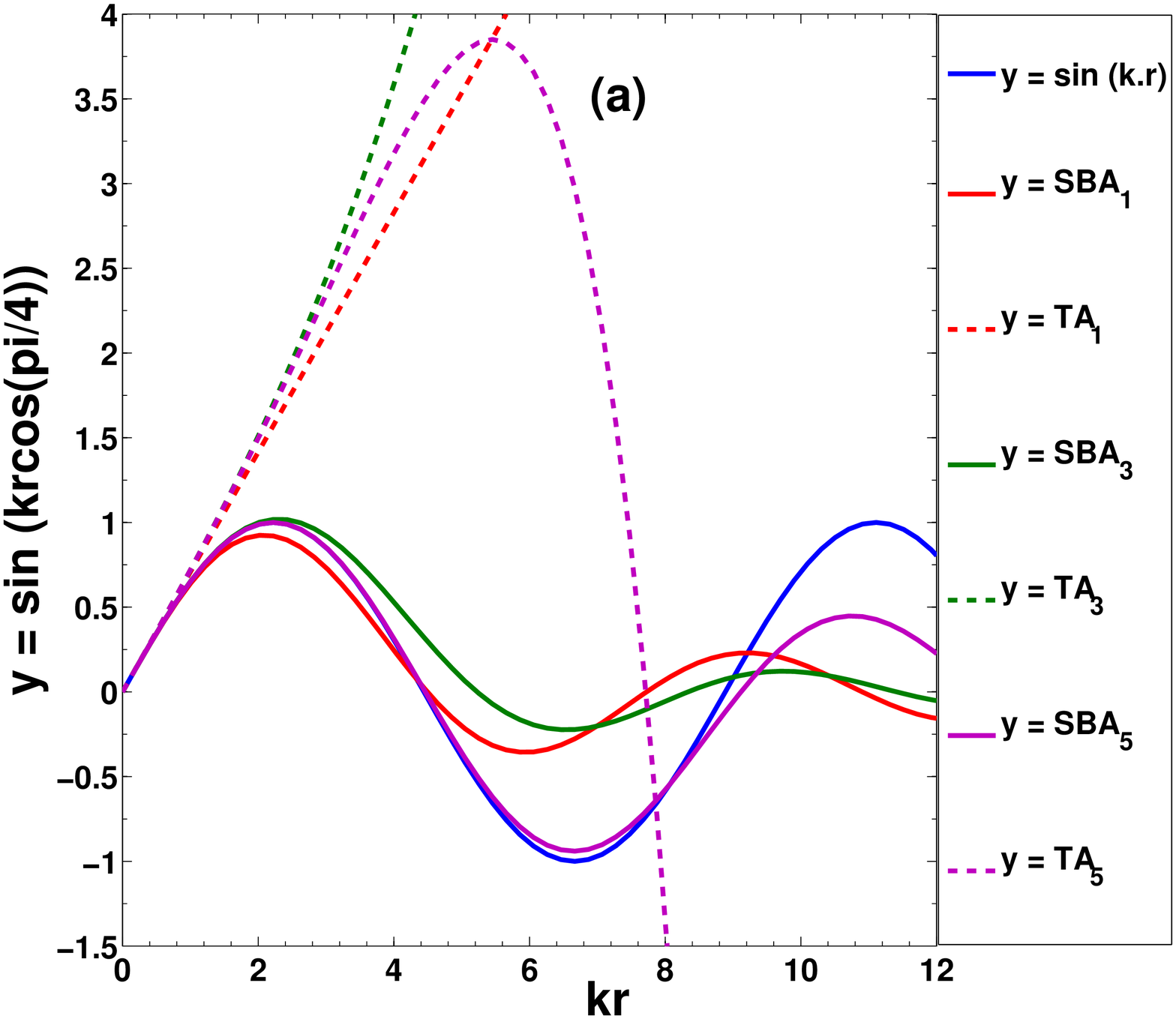}
    \includegraphics*[width= 0.49\linewidth]{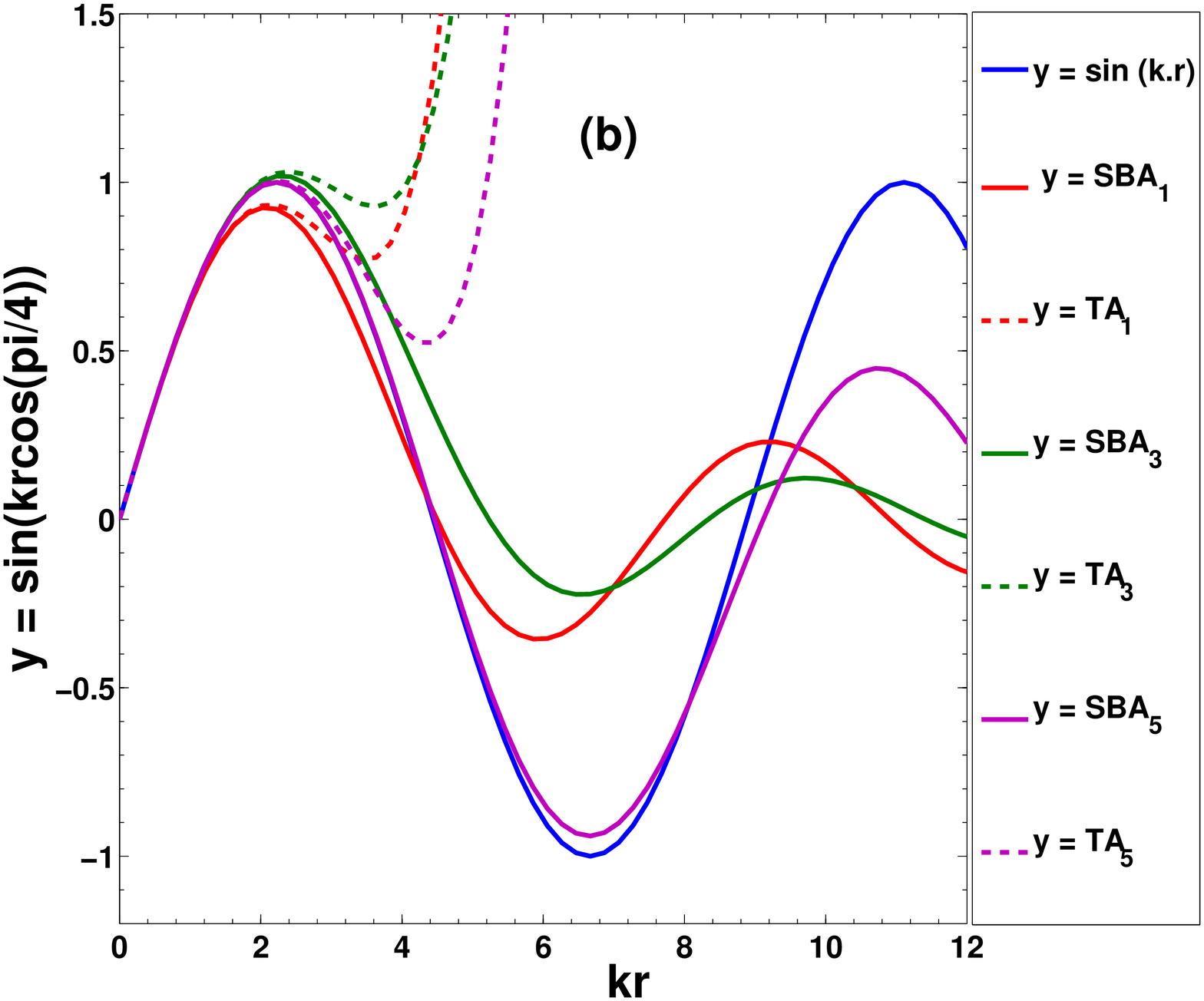}
    
    \caption{(Color online) Same as figure \ref{fig:ca} but for the sine function ${y =\mathrm{sin} (\mathbf{k}\cdot \mathbf{r})}$. }
  \label{fig:sa}
\end{figure}
  
 Figures \ref{fig:ca} and \ref{fig:sa} show the approximation of ${\cos (\mathbf{k}\cdot \mathbf{r})}$ and ${\mathrm{sin} (\mathbf{k}\cdot \mathbf{r})}$ 
 respectively using the regular spherical Bessel functions in comparison to the ones constructed from the Taylor approximated spherical Bessel functions of
 a (a) one-term ($0^{\mathrm{th}}$ degree) and (b) three-term ($2^{\mathrm{nd}}$ degree) approximation of equation (\ref{eq:jlh1}).  The angle between vectors $\mathbf{k}$ and $\mathbf{r}$ is arbitrarily chosen and here we report only the case where the angle is $\pi/4$ radians.
 
  In the figure legends we have used the functions $\mathrm{SBA}_N$ and $\mathrm{TA}_N$
     \begin{equation}
      \begin{split}
      \mathrm{SBA}_N(kr) &=  \sum_{l=even/odd}^N i^{l'} \alpha_l(r,t) j_l(kr)\,(2l+1)\mathrm{P}_l(\cos (\theta))  \\
      \mathrm{TA}_N(kr) &= \sum_{l = even/odd}^N i^{l'}\alpha_l(r,t) \tilde{j}_l^{(l_{\mathrm{max}})}(kr)\,(2l+1)\mathrm{P}_l(\cos (\theta)) \label{eq:sum1}
      \end{split}
     \end{equation}
 to denote the order of approximation of the trigonometric functions according to equation (\ref{eq:trig}) and to distinguish between the use of the regular spherical Bessel functions in the Rayleigh approximation (SBA) and the approximated spherical Bessel functions in the Taylor approximation (TA)  respectively. In the approximation of ${\cos (\mathbf{k}\cdot \mathbf{r})}$ only even terms are added with $l' = l$, while for $\mathrm{sin} (\mathbf{k}\cdot \mathbf{r})$ only odd terms are in the summation with $l' = l-1$. Both even and odd terms are used in the ionization probabilities calculations. The integer $N$ defines the upper limit in the summation and it is related to the order $l$ of spherical Bessel functions and the maximum multipole-order $l_{\mathrm{max}}$ used in the approximation by $l\leq N \leq l_{\mathrm{max}}$. In the case examples discussed in this paper, TA$_N$ and SBA$_N$ also specify the inclusion of matrix element evaluated up to a particular order and in this case all the contributions, regardless of even or odd, would be part of this sum. The value of the factor $\alpha_l$ in equation (\ref{eq:sum1}) is unity when approximating the trigonometric functions but for the ionization probabilities, it is determined by the laser as well as the system parameters making it quite complex.  

\begin{figure}[H]
  \center
    \includegraphics*[width= 0.9\linewidth]{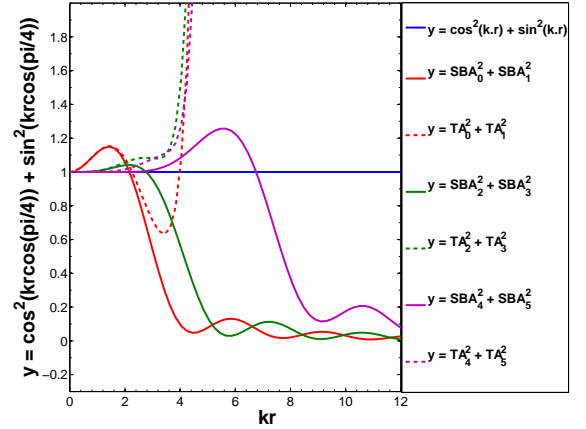}
    \caption{(Color online) Same as figure \ref{fig:ca}b but for the function ${y =\cos^2 (\mathbf{k}\cdot \mathbf{r})+\mathrm{sin}^2 (\mathbf{k}\cdot \mathbf{r})}$.}
   \label{fig:vc}
\end{figure}

 From figures \ref{fig:ca} and \ref{fig:sa}, it can be seen that the regular spherical Bessel functions used in the Rayleigh plane-wave expansion yield a better approximation of the trigonometric functions as compared to the Taylor approximated functions and the validity regime of the Taylor approximation increases with degree of summation. In figure \ref{fig:vc}, we plot further the stringent validity condition, ${\cos^2 (\mathbf{k}\cdot \mathbf{r}) + 
 \mathrm{sin}^2 (\mathbf{k}\cdot \mathbf{r})=1 }$, that need to 
 be satisfied by the various orders of the multipole expansion approximation. This helps us to visualize the spatial extent for which the various multipole-orders of each approximation satisfy the validity condition. The inclusion of higher multipole-order terms in both expansions leads to a better approximation of the spatial retardation term to a larger spatial extent as expected.
 
 It is evident from figure \ref{fig:vc} that the spherical Bessel function approximation yields a wider validity range in comparison to the Taylor approximated functions. 
 It is important to note that even though the usual dipole approximation where ${\cos (\mathbf{k}\cdot \mathbf{r})}$ is approximated to unity is quite basic,
 it fully satisfies the constraint requirement of the validity condition. This makes it quite stable against any
 breakdown in comparison to other orders of the Taylor approximation although its inherent inaccuracies remain embedded in the photoelectron energy spectrum or the photoelectron angular distribution. The point of breakdown could be conspicuously manifested in either the energy spectrum or the angular distribution. The breakdown of perturbative methods is well known and has been discussed in non-linear dynamics \cite{Strogatz2000}.
 
 In equation (\ref{eq:jlh1}), we have shown that the spherical Bessel functions can be obtained perturbatively through a power series summation of some polynomials derived from the Taylor expansion of the $\exp (\mathbf{k}\cdot \mathbf{r})$. The relation shows that the multipole transition matrices generated using the spherical Bessel functions may be equal to those generated from an infinite Taylor expansion. We probe the summation further in order to determine the optimal number of terms or rather the maximum degree of the Taylor expansion that would yield a near-exact value of the spherical Bessel generated transition matrices of the corresponding order. We used the lowest-order perturbation theory (LOPT) as defined by the Fermi-Golden rules to compare the zeroth-order multipole as well as the zeroth- plus first-order multipole energy resolved transition probabilities in both spherical Bessel functions SBA$_N$ and various degrees of the Taylor expansion TA$_N^{(n)}$. We observed that $\sim 1$ nm wavelengths are ideal for illustrating the convergence between the two approaches. For wavelengths less than the $1$ nm, it is found that the Taylor approximation is likely to break down as result of violation of the $kr \ll 1$ requirement. Other than the wavelength, the laser and numerical parameters used in the calculations were ${E_0= 50}$ {a.u.} peak electric field strength, ${r_{\mathrm{max}} = 200}$, $600$ B splines of order $10$, and a geometric knot sequence. In our calculations, we have focussed only on the $\mathbf{A} \cdot \mathbf{p}$ transitions. 
 
  \begin{figure}[H]
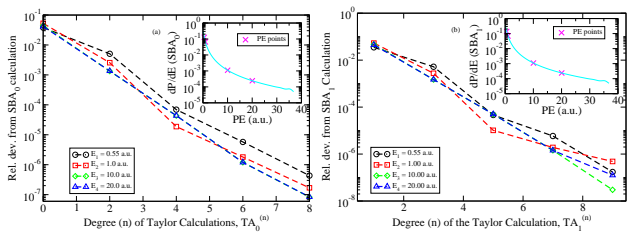

  \center
    \includegraphics*[width= 0.47\linewidth]{h200_600e100L05_1nm_TABD_C01nm_15osc_50pt0_0x_1S_V_ceaD_FGCONV_0p5-20au.eps}
    \includegraphics*[width= 0.47\linewidth]{h200_600e100L05_1nm_TABD_C01nm_15osc_50pt0_1x_1S_V_ceaD_FGCONV_0p5-20au.eps}
    
    \caption{(Color online) Convergence of different degrees of (a) zeroth order Taylor, TA$_0^{(n)}$, (b) zeroth plus first order Taylor, TA$_1^{(n)}$, energy resolved ionization probabilities calculated perturbatively relative to the corresponding SBA$_0$ and SBA$_1$ calculations respectively using the Fermi-golden rules. Insets: The extracted SBA benchmark energy resolved photoionization probabilities specified by the crosses.}
  \label{fig:fg}
\end{figure}
 
 Figure (\ref{fig:fg}) shows the relative deviations of the $n^{th}$ degree Taylor TA$_N^{(n)}$ multipole energy resolved probabilities from the corresponding SBA$_N$ reference calculations. The relative deviation in this case is calculated by getting the absolute difference between the TA$_N^{(n)}$ and SBA$_N$ calculation and then dividing the result by the   SBA$_N$ calculation at a fixed photoelectron energy. In the insets, the four extracted energy points and their probabilities are indicated by the crosses. It is observed that TA$_N^{(n)}$ multipole probabilities approach the SBA$_N$ probabilities as the degree $n$ of Taylor expansion increases. It can be seen that the probabilities converge at slightly different rates for different points of energy but the overall trend is the same. However, we did observe that beyond $n=9$, there is no further convergence of probabilities and divergence  sets in. The cause of the divergence could not be exactly resolved. Further investigation  on the cause of these divergence is absolutely necessary.
 
 As a non-perturbative example, we consider the multiphoton above-threshold-ionization of a hydrogen atom initially in ground state irradiated by superintense free-electron XUV or x-ray
 laser fields. First, we chose the radiation wavelength of $9.11$ nm and the field intensity of $50 \,\mathrm{a.u.}$ consistent with the results of reference \cite{Zhongyuan2013}. Similar to the perturbative calculations, we have implemented the multipole $\mathbf{A} \cdot \mathbf{p}$ transitions from the zeroth- to the third-order using both Taylor (TA) and spherical Bessel function (SBA) expansions. Each higher-order calculation includes the contribution of all the previous orders. Our goal was to clearly show the disparities, arising from the different expansion approaches, basically in the multipole photoelectron energy spectra evaluated variationally without incorporating too many interactions and also to compare their respective computational advantages.  
 
 \begin{figure}[H]
  \center
    \includegraphics*[width= 0.9\linewidth]{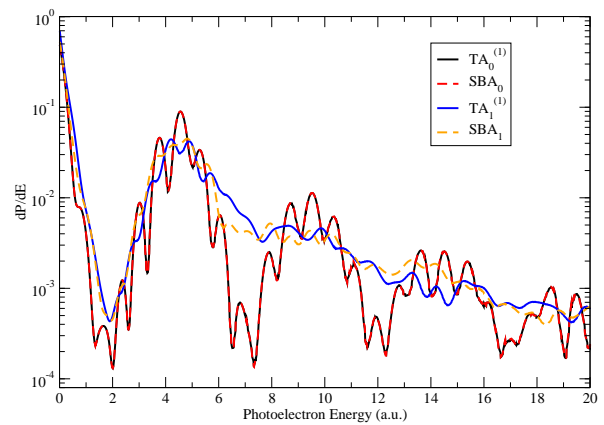}    
    \caption{(Color online) Photoelectron energy spectrum of multiphoton ionization of hydrogen atom using $9.11$ nm radiation field with a peak electric field amplitude of ${E_0 = 50 \; \mathrm{a.u.}}$ corresponding to an intensity of ${8.77\times10^{19} \mathrm{Wcm}^{-2} }$ and Reiss intensity scaling parameter ${z = 4.9956}$. Solid lines correspond to $N^{\mathrm{th}}$ order Taylor expansion (TA$_N^{(1)}$) in accordance with equation (\ref{eq:jlh1}) and consistent with the order of Taylor expansion series. Broken lines to spherical Bessel expansion (SBA$_N$)of the corresponding orders .
     Black: TA$_0^{(3)}$,  red: SBA$_0$, blue: TA$_1^{(3)}$, orange: SBA$_1$. } 
  \label{fig:pe0}
\end{figure}
 
\begin{figure}[H]
  \center
    \includegraphics*[width= 0.9\linewidth]{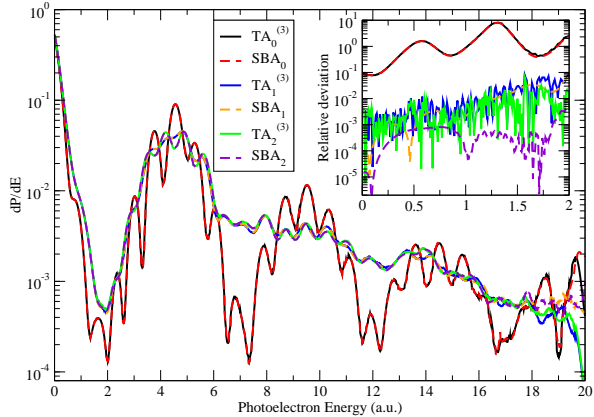}    
    \caption{(Color online) Same laser parameters as figure (\ref{fig:pe0}) but with additional higher order multipole probability distribution.  Black: TA$_0^{(3)}$,  red: SBA$_0$, blue: TA$_1^{(3)}$, orange: SBA$_1$, dark green:TA$_2^{(3)}$,and violet: SBA$_2$. Inset: The corresponding succesive relative deviation. } 
  \label{fig:pe1}
\end{figure}

The photoelectron energy spectrum evaluated using 
 both spherical Bessel and Taylor expansion for the multipole transitions up to the first- and third-order are shown in figures (\ref{fig:pe0}) and (\ref{fig:pe1}) respectively, with focus in the $\mathbf{A} \cdot \mathbf{p}$ interactions only. In our calculations,  we obtained convergence for the four-photon ionization above-the-threshold ionization peaks corresponding to the $9.11$ nm wavelength \cite{Zhongyuan2013} using ${L_{max} = M_{max} = 5}$, Box radius $= 200$ au, B splines $=600$, velocity gauge, cut-off energy $= 20$ a.u., tolerance $=10^{-10}$, and a fully implicit time propagation scheme. In evaluating the transition matrices corresponding to this figure, we used the Taylor expansion of the first two terms in figure (\ref{fig:pe0}) from which we derived the Taylor approximated spherical Bessel functions  $\tilde{j}_0^{(1)}(kr)$, $\tilde{j}_1^{(1)}(kr)$, and the first four terms (\ref{fig:pe1}) from which we derived the Taylor approximated spherical Bessel functions  $\tilde{j}_0^{(3)}(kr)$, $\tilde{j}_1^{(3)}(kr)$, $\tilde{j}_2^{(3)}(kr)$, and $\tilde{j}_3^{(3)}(kr)$. These would then be used in the evaluation of the $N^{\mathrm{th}}$ order Taylor approximated multipole transition matrices (TA$_N^{(l_{max})}$). The results are compared with the corresponding SBA$_N$ calculations. We find out that for the photoelectron energy window specified in the figure, the second order multipole terms (TA$_2$ and SBA$_2$) are already sufficient for perfect convergence. The comparison between TA$_N$ and SBA$_N$ spectra can be seen to be very good for the chosen level of Taylor expansion. However, when $l_{max}<3$ is used as the maximum order of Taylor expansion as can be evidently seen in figure (\ref{fig:pe0},) we find that the agreement between the two approaches  for the same laser and numerical parameters is not very good except for the dipole approximation. This therefore confirms that higher orders of Taylor expansion are indeed necessary to obtain the near exact accuracy with spherical Bessel functions as already shown in theory. In the figure caption, we have included the Reiss intensity scaling parameter which shows the ratio of the ponderomotive energy to the photon energy and as a consequence it measures the degree of applicability of the perturbation theory. 
  
\begin{figure}[H]
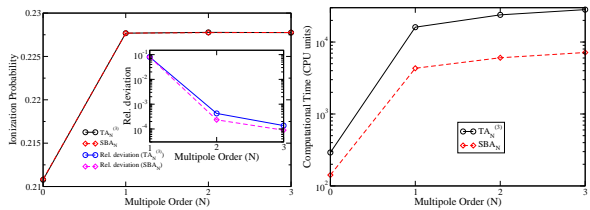

  \center
    \includegraphics*[width= 0.44\linewidth]{h200_600e100L05_9pt11nm_TAnSBABD_C911_10osc_50_0123x_1S_V_cFaA_IonYield2_ok.eps}   
    \includegraphics*[width= 0.44\linewidth]{h200_600e100L05_9pt11nm_TAnSBABD_C911_10osc_50_0123x_1S_V_cFaA_CTC_ok.eps}   
    \caption{(Color online) (a) Total ionization probability and relative deviation (inset) of the ionization yield from the previous order interactions, and (b) computational run-time as a function of the multipole order of the retardation expansions in Taylor or in spherical Bessel function series. The numerical and laser parameters are similar to those used in figure (\ref{fig:pe1}).} 
  \label{fig:cpc1}
\end{figure}

 In general, we observe that non-dipole corrections may provide significant disparities in the spectra depending on the method of expanding the retardation term. The effect of including the quadrupole and the octupole $\mathbf{A} \cdot \mathbf{p}$ terms as the only corrections to the dipole approximation already shows some significant transformation of the ATI structures predicted by the dipole approximation similar to the observations in the literature data \cite{Zhongyuan2013}. The side bands are apparently levelled out and the second-photon peak flattened by this partial inclusion of non-dipole effects. The magnitude of the non-dipole effects on the photoelectron energy spectrum is observed to increase with the photoelectron energies but the general non-dipole induced  interference of spectrum is spread at all photoelectron energies. In the PE spectrum, it can be observed that tiny higher multipole-order effects also manifest slightly at very low photoelectron energies. As photoelectron  energies increase, the disparity between various orders become more pronounced necessitating the need for higher-order corrections. The dominant non-dipole effects can be attributed to the quadrupole transitions and the higher multipole-order interactions only produce tiny modulations to the quadrupole effects.  This is further confirmed in figure (\ref{fig:cpc1} a)  showing relative deviations of the total ionization yield from the lowest-order spectra with the biggest deviation emanating from the $1^{\mathrm{st}}$ order correction.  The discrepancy between the methods of expanding the retardation is also clearly visible and can not be perfectly resolved  even by increasing the multipole orders of expansion. Our objective of introducing the higher-order transitions was to actually test convergence between the spectra by increasing the order of the multipole expansion in this two-photon resonance regime. We obtain near-perfect convergence between the two expansions only in the  $3^{\mathrm{rd}}$ multipole-order approximation but the relative deviation in ionization probability from the previous order spectra shown in figure (\ref{fig:cpc1} a) suggests that there always exist a finite uncertainty between the corresponding orders in the total ionization yield. This relative deviation is evaluated by taking the difference in ionization probability between successive orders divided by the lower order ionization probability. It can be observed that the SBA relative deviations are smooth varying functions while those from the TA calculations are saw-toothed varying functions. At the highest multipole-order terms considered, the SBA relative deviation can be seen to be much smaller than the TA relative deviations. We have already seen from the perturbative treatment that very high orders of Taylor expansion is required to resolve its disparity with SBA calculations. 

\begin{figure}[H]
  \center
    \includegraphics*[width= 0.9\linewidth]{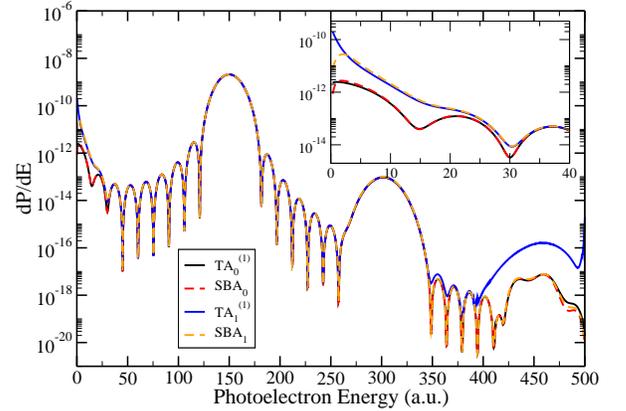}   
    \caption{(Color online) Same as figure (\ref{fig:pe1}) but for a $0.3$ nm wavelength and Reiss intensity scaling parameter $z = 1.784 \times 10^{-4}$. Inset: Lower photoelectron energy regime showing some dicrepancy between the TA$_N^{(1)}$ and SBA$_N$ approximations and the manifestation of first-order correction effects. } 
  \label{fig:pe2}
\end{figure}

\begin{figure}[H]
  \center
    \includegraphics*[width= 0.85\linewidth]{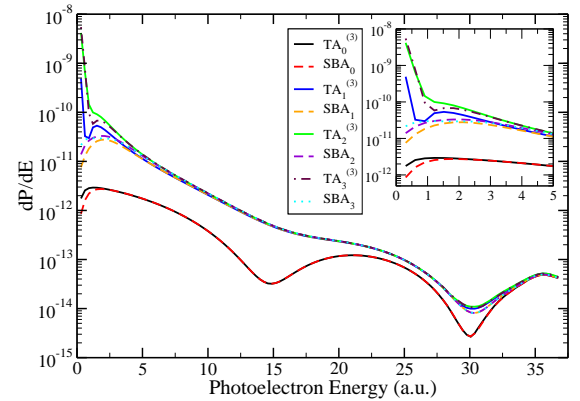}   
    \caption{(Color online) Same as figure (\ref{fig:pe2}) but including the effect of higher order multipoles up to the order of hexadecapole contribution in the lower photoelectron energy region. We increased the order of multipole expansion to include TA$_3$ (cyan) and SBA$_3$ (light green). } 
  \label{fig:pe3}
\end{figure}

 Figure (\ref{fig:pe2}) shows a similar comparison of the photoelectron spectrum at a much shorter wavelegth of $0.3$ nm. The spectrum shows a series of side bands, which are signatures of dynamic Stark effect \cite{Demekhin2012, Chuan2013}, and three multi-photon resonance peaks are also visible. The figure was generated using similar numerical parameters except the number of B splines which were increased to $5000$ in order to be able to treat the higher photo-electron energies upto the range specified. The discrepancy between the  TA$_N$ and SBA$_N$ spectra are observable at lowest-order approximation  and also at the nearthreshold photoelectron energies. The first-order beyond the dipole correction effects manifesting at the lower photoelectron energies are supported by both approaches even though the estimated magnitude differs. At highest photoelectron energies, there appears to be some slight spatial effect in SBA$_1$ prediction but the TA$_1$ predicted non-dipole effect is quite huge. Considering that the multipole expansion in spherical Bessels functions give the ideal contribution from mathematical perspective, one can argue that the non-dipole structure observed when Taylor expansion is used seems to be more of an artefact. This reasoning is quite logical basing on the fact that Taylor expansion is valid only when ${2\pi/ \lambda \ll 1}$. For $0.3$ nm, ${2\pi/ \lambda \sim 1}$ which definitely contravenes the validity condition. This may suggest that the Taylor approximation breaks down in this regime  and its use would only yield non-physical results. In the inset in figure (\ref{fig:pe2}) we zoom into the low-energy window showing some non-dipole effects. The energy range shown is within the single-photon ionization regime. We probed the disparity between the two alternative expansions further by considering the contribution of higher-order multipoles in the lower photoelectron energy regime where the difference is larger. Figure (\ref{fig:pe3}) shows the spectra evaluated up to the $3^{\mathrm{rd}}$ multipole-order contributions. Although these laser parameters make the interactions to be classified under the dipole oasis \cite{Ludwig2014} by the scaling laws, it is surprising that the role of the additional higher multipole-order effects are still significant in this energy regime.  In this figure, the discrepancy between TA and SBA at very low photoelectron energies persists despite the involvement of many orders of the multipole expansion. It can also be recognized that higher multipole-orders make a significant contribution in this near-threshold region, and more higher multipole-orders are still necessary for perfect convergence of the spectra.  A small bending in the TA$_0$ and TA$_1$ spectra in the near-threshold region is noticeable  because of the Taylor approximated spherical Bessel function ($\tilde{j}_{0}^{(3)}(kr)$ and $\tilde{j}_{1}^{(3)}(kr)$) used incorporates the $1^{\mathrm{st}}$ higher degree corrections. This makes the Taylor approximated spectra to be slightly like the spherical Bessel functions generated spectra. In the perturbative treatment, we have already shown that several higher degree corrections may be necessary, even in the regime where Taylor approximation does not breakdown,  for equivalent treatment in both TA and SBA expansions.
 
 In general, there appears to be some numerical gain with the use of spherical Bessel functions compared to Taylor approximation in terms of the computational run-time for the time propagation.   The computation run time of $9.11$ nm wavelength laser pulse interaction calculated for up to $N^{\mathrm{th}}$ order approximation in Taylor and spherical Bessel functions respectively consistent with the numerical parameters used in our calculations is illustrated in figure(\ref{fig:cpc1}b). Beyond the dipole, the  computational run-time for time propagation dynamics shows logarithmic time dependence as the multipole orders increase. It can be seen that the SBA calculations are more efficient in time. The gain in time when SBA is used was observed to be even more significant when shorter wavelengths ($\sim 0.3$ nm and below) are used.  A detailed analysis of the non-dipole effects including the effects of the $\mathbf{A} \cdot \mathbf{A}$ interactions will be discussed in a subsequent publication.

\section{Conclusion}
\label{sec:conc}
We have compared the multipole expansion of the retardation term $e^{i\mathbf{k} \cdot \mathbf{r}}$ using both the
Taylor approximation and the spherical Bessel function approximation. We find that using the spherical Bessel functions in
increasing orders accurately tracks the retardation term to better accuracy than the Taylor series. Moreover, Taylor approximation is shown to converge to the  Rayleigh multipole-order expansion using spherical Bessel
functions in the limit $kr \rightarrow 0$ although at large $kr$ there are larger deviations between the two approximations. Using higher multipole-orders as well as higher degrees of the Taylor expansions reduce their disparity with the Rayleigh plane-wave multipole expansion.
From the theory, one can conclude that the Rayleigh plane-wave multipole expansion provides the ideal contribution
of any given order of the multipole expansion. In the example calculations of the ionization probability distributions, we show that the zero-order terms compare very well in both expansions and that essentially means that the dipole approximation would be justified when the importance of the non-dipole corrections is not necessary. When non-dipole effects are of interest, the discrepancy manifesting between the expansions show that some amount of discrepancy will always exist in the calculated spectrum depending on the method of expanding the retardation term. For the case considered in this study, the discrepancy accounts for the role of higher multipole-order interactions which are not present in the Taylor expansion.  For very short wavelengths and at very low photoelectron energies, the structure of the photoelectron distibution spectrum varies slightly depending on the method of expansion employed signifying that the inclusion of many higher-order non-dipole effects may be crucial to predict correctly the photoelectron spectrum in this energy regime for extremely short wavelengths. It may be important to note that the validity condition of Taylor approximations requiring the correction terms to be far much less than unity fails in the very short wavelength domain. This could explain the reason why the Taylor corrections are badly behaved in the case of $0.3$ nm shown in figure (\ref{fig:pe3}) as compared to the case of $9.11$ nm shown in figure (\ref{fig:pe1}). Using SBA is observed to be practically more efficient in terms of the computational time and the ideal accuracy for any given order of the multipole expansion.

\bibliographystyle{apsrev}
\bibliography{/home/eric/Inworks/Literature}

\end{document}